\documentclass[aps,twocolumn,superscriptaddress,floatfix,preprintnumbers,nofootinbib]{revtex4-2}
\usepackage[utf8]{inputenc}
\usepackage{amsmath}
\usepackage{amssymb}
\usepackage{multirow}

\usepackage{enumerate}
\usepackage{amsmath}
\usepackage{tikz}
\usepackage[compat=1.1.0]{tikz-feynman}
\usepackage{feynmf}
\usepackage{slashed}
\usepackage{braket}
\usepackage[toc,page]{appendix}
\usepackage{url}
\usepackage{natbib}
\usepackage{graphicx}
\usepackage[pdftex,bookmarks,linktocpage,pdfpagelabels,plainpages=false,hyperfigures,linkcolor=blue,citecolor=blue]{hyperref}
\hypersetup{colorlinks=true}

\usepackage{mathrsfs,amssymb}  
\usepackage{cancel}
\usepackage[normalem]{ulem}

\begin{document}

\begin{flushright}
\preprint{MI-TH-2110} 
\end{flushright}

\author{P.~S.~Bhupal Dev}
\email{bdev@wustl.edu}
\affiliation{Department of Physics and McDonnell Center for the Space Sciences, Washington University, St. Louis, MO 63130, USA}

\author{Doojin~Kim}
\email{doojin.kim@tamu.edu}
\affiliation{Mitchell Institute for Fundamental Physics and Astronomy,
Department of Physics and Astronomy, Texas A\&M University, College Station, TX 77843, USA}

\author{Kuver Sinha}
\email{kuver.sinha@ou.edu}
\affiliation{Department of Physics and Astronomy, University of Oklahoma, Norman, OK 73019, USA}

\author{Yongchao Zhang}
\email{zhangyongchao@seu.edu.cn}
\affiliation{School of Physics, Southeast University, Nanjing 211189, China}

\title{New interference effects from light gauge bosons in neutrino-electron scattering
}

\begin{abstract}
We point out that light gauge boson mediators could induce new interference effects  in neutrino-electron scattering that can be used to enhance the sensitivity of neutrino-flavor-selective high-intensity neutrino experiments, such as DUNE. We particularly emphasize on a destructive interference effect, leading to a deficit between the  Standard Model expectation and the experimental measurement of the differential cross sections, which is prominent only in either the neutrino or the antineutrino mode, depending on the mediator couplings. Therefore, the individual neutrino (or antineutrino) mode  could allow for sensitivity reaches superior to the combined analysis, and moreover, could distinguish between different types of gauge boson mediators.   
\end{abstract}

\maketitle

\section{Introduction} 
Scattering experiments involving neutrino beams have played a crucial role in the development and precision studies of the Standard Model (SM) of particle physics~\cite{Conrad:1997ne}. Since neutrinos only couple to the SM particles through the charged-current and neutral-current weak interactions, which are well determined, neutrino scattering  measurements can be used as a precision probe of the SM. In particular, since neutrino-electron elastic scattering cross-sections in the SM are predicted to high accuracy~\cite{Tomalak:2019ibg, Miranda:2021mqb}, any significant deviation between measurements and the SM expectation would indicate new physics. 

New interactions in the neutrino sector that could potentially lead to  detectable new physics signatures in current and future neutrino-electron scattering experiments can be motivated from various angles. For instance, the observed neutrino oscillation data require the existence of some new physics involving the neutrino sector, which necessarily imply new neutrino interactions at either tree- or loop-level. Similarly, the observed relic density of dark matter in the universe can be explained by invoking various portals connecting the dark/hidden sector to the SM, which often involve new neutrino interactions. These general new physics arguments have inspired several phenomenological studies analyzing the effects of new interactions between neutrinos and electrons, either in a simplified model framework~\cite{Boehm:2004uq,Harnik:2012ni,Chiang:2012ww, Billard:2014yka,  Bilmis:2015lja, Farzan:2015doa, Cerdeno:2016sfi, Deniz:2017zok, Ge:2017mcq, Lindner:2018kjo, Wise:2018rnb, Ballett:2019xoj, Berryman:2019dme,  Miranda:2020zji, Abe:2020nwr, Amaral:2020tga,Bally:2020yid,Ibe:2020dly,Amaral:2021rzw, Dev:2021qjj} or in an effective field theory set-up~\cite{Barranco:2007ej, Bolanos:2008km,Sevda:2016otj, Sobkow:2016zsg, Rodejohann:2017vup, Falkowski:2018dmy,Link:2019pbm, Khan:2019jvr, Chen:2021uuw, deSalas:2021aeh}. Here we will consider one such example with a new light gauge boson interacting with neutrinos and electrons. This can arise naturally if one of the accidental global symmetries of the SM is gauged~\cite{He:1991qd}.\footnote{Note that the gauged subgroups of lepton number, $U(1)_{L_\alpha-L_\beta}$ (with $\alpha,\beta\in \{e,\mu,\tau\}$ and $\alpha\neq \beta$), are anomaly-free without any additional particles, whereas $U(1)_{B-L}$ can be made anomaly-free if right-handed neutrinos are introduced.} 

In most of the well-motivated and allowed regions of their parameter space, such light gauge bosons feebly interact with SM particles. Therefore,  high-intensity beam-based experiments are expected to have great sensitivity to the associated signals. 
In particular, current and future neutrino-beam  facilities, such as the NuMI neutrino beam~\cite{Adamson:2015dkw}, carry great capabilities of probing light gauge boson mediators. 
In the presence of such a mediator, neutrinos can scatter off the target material in the detector via a $t$-channel exchange of the mediator, on top of the SM interactions. 
Therefore, any excess or deficit beyond SM predictions can be interpreted as evidence of such light mediators. 

In this paper, we carefully delve into phenomenological impacts of the interference effect.  
We point out that destructive interference due to a light vector mediator may significantly diminish the SM expectation,\footnote{For a related discussion, see Refs.~\cite{Ballett:2019xoj,Amaral:2020tga, Amaral:2021rzw}.} hence a {\it deficit} may make manifest the existence of the light gauge bosons, especially in {\it neutrino-flavor-selective experiments} like DUNE~\cite{AbedAbud:2021hpb} via beam focusing horns. 
We further, for the first time, find that the destructive interference effect is prominent {\it only} in either the neutrino mode or the antineutrino mode, depending on the mediator type. Thus the {\it individual analyses} of neutrino and antineutrino modes could allow for sensitivity reaches superior to the combined analysis.
To demonstrate our findings, we take the $B-L$ and $L_e-L_\mu$ gauge bosons as case studies, and analyze their searches in neutrino-electron scattering  at the DUNE near-detector~\cite{AbedAbud:2021hpb} and at JSNS$^2$~\cite{Ajimura:2017fld} experiments as concrete examples of flavor-selective and non-flavor-selective experiments, respectively.  

The rest of the paper is organized as follows. In Sect.~\ref{sec:form} we describe and extend the formalism of the scattering between a neutrino and an electron, taking into account contributions of a light gauge boson including the interference effect. The benchmark experiments to which our main idea is applicable are briefly described in Sect.~\ref{sec:exp}, followed by the strategy of our data analyses in Sect.~\ref{sec:analysis}. The main results including the sensitivity reaches of the two example cases will appear in Sect.~\ref{sec:results}. Sections~\ref{sec:discussion} and \ref{sec:conclusion} are reserved for further discussion and conclusions, respectively. The formalism in the case of general $Z'$ gauge boson interactions is explained in the appendix.

\section{Formalism \label{sec:form}}
We briefly discuss the scattering of neutrinos off an electron in the presence of a light gauge boson under which the electron and neutrinos (not necessarily all flavors) are non-trivially charged. 
The relevant part of the interaction Lagrangian can be generically expressed as follows:
\begin{equation}
    - \mathcal{L} \supset g_V Q_{\ell} V_\mu \bar{\ell}\gamma^\mu \ell + g_V Q_{\nu_\ell} V_\mu \bar{\nu}_\ell\gamma^\mu \nu_\ell\,,
\end{equation}
where $g_V$ denotes the coupling of the light gauge boson $V$ and $Q_\ell$, $Q_{\nu_\ell}$ parameterize the gauge charges of SM leptons and neutrinos, respectively. 

\begin{figure}[t]
\centering
\includegraphics[width=8.4cm]{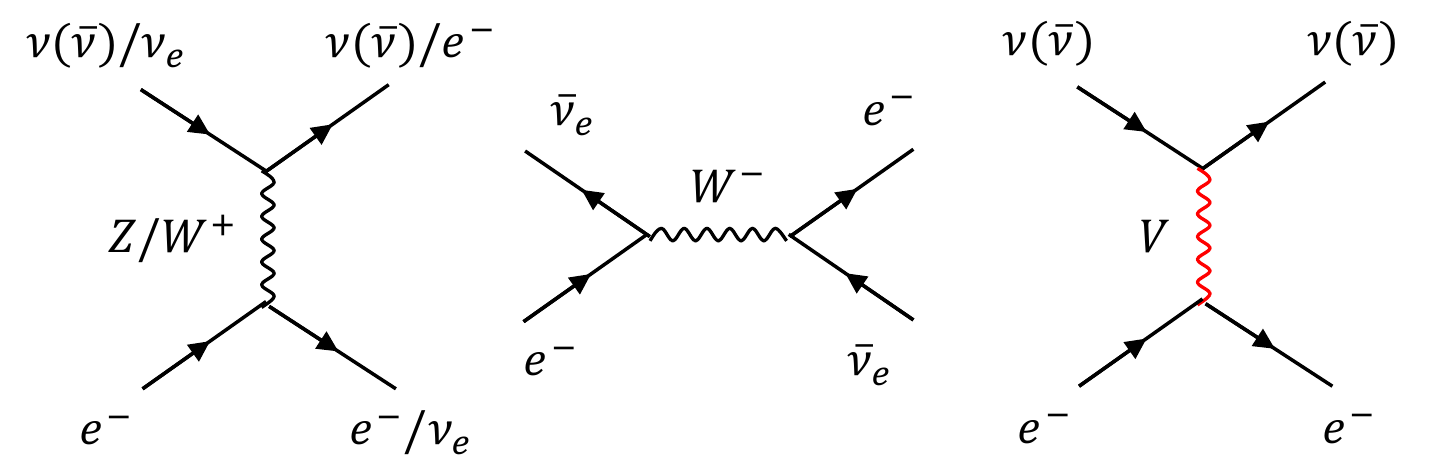}
\caption{Feynman diagrams of the SM processes (first and second) and the light gauge boson contribution (third) for the neutrino-electron elastic scattering. For the cases with antineutrinos in the parentheses, the arrow directions are assumed opposite. }
\label{fig:diagram}
\end{figure}

In the presence of the above interaction terms, a neutrino can scatter off an electron through a $t$-channel exchange of $V$ (third diagram in Fig.~\ref{fig:diagram}), in addition to the SM processes (first and second diagrams in Fig.~\ref{fig:diagram}). 
Since they all involve the same incoming and outgoing states, their amplitudes interfere with one another quantum-mechanically. 
The overall cross-section $\sigma$ is given by the sum of the pure SM contribution $\sigma_{\rm SM}$, the pure $V$ contribution $\sigma_V$, and the interference $\sigma_{\rm int}$. We then express the differential cross-section in the energy of the recoiling electron $E_e$ as
\begin{equation}
    \frac{d\sigma}{dE_e}=\frac{d\sigma_{\rm SM}}{dE_e}+\frac{d\sigma_V}{dE_e}+\frac{d\sigma_{\rm int}}{dE_e}\,.
\end{equation}
The three terms on the right-hand side are given by
\begin{eqnarray}
    \frac{d\sigma_{\rm SM}}{dE_e} &=& \frac{2G_F^2m_e}{\pi E_\nu^2}\left\{c_1^2E_\nu^2+c_2^2(E_\nu-E_e)^2-c_1c_2m_e E_e \right\}, \nonumber \\
    && \label{eq:smcont} \\
    \frac{d\sigma_V}{dE_e} &=&\frac{Q_{\nu_\ell}^2Q_e^2g_V^4m_e}{4\pi E_\nu^2}\frac{\left\{2(E_\nu-E_e)E_\nu+(E_e-m_e)E_e \right\}}{(2m_eE_e+m_V^2)^2}, \nonumber \\
    && \label{eq:zpcont}\\
    \frac{d\sigma_{\rm int}}{dE_e} &=& \frac{Q_{\nu_\ell}Q_eg_V^2G_Fm_e}{2\sqrt{2}E_\nu^2\pi(2m_eE_e+m_V^2)} \nonumber \\
    &\times& \left\{c_3(2E_\nu^2-m_eE_e)+c_4 2(2E_\nu-E_e)E_e \right. \nonumber \\
    &+& \left. 4s_W^2\left[2(E_\nu-E_e)E_\nu+(E_e-m_e)E_e\right]\right\}, \label{eq:intcont}
\end{eqnarray}
where $G_F$, $E_\nu$, $m_e$, and $m_V$ denote the Fermi constant, the incoming neutrino energy, the electron mass, and the $V$ gauge boson mass, respectively. 
The four flavor-dependent coefficients, $c_1$ through $c_4$, are summarized in Table~\ref{tab:coeff}.

\begin{table}[t]
    \centering
    \begin{tabular}{c|c | c| c| c}
    \hline \hline
         ~~~Flavor~~~ & ~~~~$c_1$~~~~ & ~~~~$c_2$~~~~ & ~~~~$c_3$~~~~ & ~~~~$c_4$~~~~  \\
         \hline 
         $\nu_e$ & $s_W^2+\frac{1}{2}$  & $s_W^2$ & $+1$ & $0$  
         \\
         $\bar{\nu}_e$ & $s_W^2$ & $s_W^2+\frac{1}{2}$ & $+1$ & $-1$  \\
         $\nu_\mu,\nu_\tau$ & $s_W^2-\frac{1}{2}$ & $s_W^2$ & $-1$  & $0$ \\
         $\bar{\nu}_\mu,\bar{\nu}_\tau$ & $s_W^2$ & $s_W^2-\frac{1}{2}$ & $-1$ & $+1$ \\
         \hline \hline 
    \end{tabular}
    \caption{A summary of the coefficients in Eqs.~\eqref{eq:smcont} through \eqref{eq:intcont}. $s_W$ defines the Weinberg angle $\theta_W$ as $s_W\equiv \sin\theta_W$.}
    \label{tab:coeff}
\end{table}

Taking an approximation of $s_W^2\approx 0.25$, one can readily find that the interference contributions of $\nu_{\mu}$ (and $\nu_\tau$ as well) and its antiparticles are reduced to
\begin{eqnarray}
    \left(\frac{d\sigma_{\rm int}}{dE_e}\right)_{\nu_{\mu}} &\propto& -Q_{\nu_{\mu}}Q_e(2E_{\nu_{\mu}}-E_e)\,,\\ 
    \left(\frac{d\sigma_{\rm int}}{dE_e}\right)_{\bar{\nu}_{\mu}} &\propto& Q_{\nu_{\mu}}Q_e(2E_{\nu_{\mu}}-E_e)\,,
\end{eqnarray}
where all the positive prefactors are dropped out. 
By construction, $E_\nu>E_e$, hence the overall sign of the interference term is determined by the sign difference of $Q_{\nu_{\mu}}$ relative to $Q_e$.
For example, in models of the $B-L$ gauge boson, all charged leptons and neutrinos have the same $U(1)$ gauge charge. Therefore, $\nu_{\mu}$ ($\bar{\nu}_\mu$) can accompany destructive (constructive) interference. 
By contrast, in models of the $L_e-L_\mu$ gauge boson, $\nu_\mu$ carries a $U(1)$ charge opposite to that of the electron. As a consequence, $\nu_{\mu}$ ($\bar{\nu}_\mu$) can give rise to constructive (destructive) interference. 
This observation is particularly useful if an experiment receives neutrino and antineutrino fluxes effectively separately.  
These different behaviors will be discussed when we show our sensitivity results for the two classes of models.

A similar exercise can be done for electron-flavored neutrinos. Unlike $\nu_{\mu,\tau}$, the net effect of their interference depends on the differential $\nu_e$ flux in energy, the electron recoil energy, and the electron mass. 
In most of the beam-based neutrino experiments, electron neutrinos do not dominate over the other flavors, hence the net interference effect can be determined in combination with the contributions of the other flavors. 
We will revisit this issue when contrasting the sensitivity reaches of our benchmark experiments. 

\section{Benchmark Experiments \label{sec:exp}} 
We apply the idea delineated in the previous section to a couple of beam-based neutrino experiments, DUNE and JSNS$^2$. 
DUNE will utilize a high-energy proton beam to produce neutrinos, while JSNS$^2$ is generating a neutrino flux with a low-energy proton beam. 
Basically, most of the neutrinos stem from the decay of mesons such as $\pi^\pm$. However, the details of obtaining neutrino fluxes differ from each other due to the difference of the beam energy, and affect the resulting signal sensitivity.  
We briefly review key features of these experiments and contrast their ways of obtaining neutrinos. 

\begin{itemize}
    \item DUNE~\cite{Abi:2020wmh}: Deep Underground Neutrino Experiment (DUNE), which will be operational in $\sim5$ years, is a multi-purpose neutrino experiment, encompassing not only neutrino phenomenology, e.g., $\delta_{CP}$ in the lepton sector, supernova neutrino observations, etc, but new physics beyond the SM including proton decays, sterile neutrinos, cosmogenic boosted dark matter etc~\cite{Abi:2020kei}. A 1.2~MW beamline will yearly deliver about $1.1\times 10^{21}$ 120-GeV protons on a graphite target. (Relatively) long-lived charged particles such as $\pi^\pm$, $K^\pm$ escape the target and enter the magnetic horn regions. 
    Depending on the current flowing in the magnets, they are focused sign-selectively, and in turn, their decay products, neutrinos are focused flavor-selectively. The neutrinos then fly to the near detector (ND) complex which is located 574~m away from the target. In the (anti)neutrino mode, the neutrino flux reaching the ND is dominated by $\nu_\mu$ ($\bar{\nu}_\mu$).  
    The ND complex consists of a liquid argon time projection chamber detector (a.k.a. ArgonCube), a gaseous argon time projection chamber detector, and a neutrino beam monitoring detector. ArgonCube with a rectangular geometry of $w\times h \times l=5~{\rm m}\times 3~{\rm m}\times 4~{\rm m}$ has a 67.5-ton fiducial mass and collects neutrino interaction data. 
    \item JSNS$^2$~\cite{Ajimura:2017fld}: The J-PARC Sterile Neutrino Search at the J-PARC Spallation Neutron Source (JSNS$^2$) experiment, which has collected data since the end of 2020, aims to probe the existence of neutrino oscillations with $\Delta m^2$ around $1~{\rm eV}^2$. 
    About $3.8\times 10^{22}$ 3-GeV protons impinge on a mercury target annually. 
    A charged pion produced by the collision of beam protons on target loses its energy, stops inside the target, and decays into a muon and a muon neutrino. The muon is not energetic, so it also stops inside the target and decays into an electron, an electron neutrino, and a muon (anti)neutrino. 
    Unlike DUNE, neutrino flavors are not selective so that similar numbers of neutrinos and antineutrinos reach the detector. Also, since neutrinos are from stationary sources (stopped pions and muons), their flux is isotropic. Once produced, neutrinos reach a gadolinium(Gd)-loaded liquid-scintillator (LS) detector which has a fiducial mass of 17 tons and is placed at a distance of 24~m from the target. The Gd-loaded LS detector is surrounded by $\sim30$ tons of unloaded LS which vetoes the background signals coming from outside. 
    The JSNS$^2$ Collaboration is now planning to deploy another detector of a 45-ton fiducial mass at a distance of 48 m from the target with the same detector technology~\cite{Ajimura:2020qni}.
\end{itemize}

\begin{figure*}[t!]
    \centering
    \includegraphics[width=8.8cm]{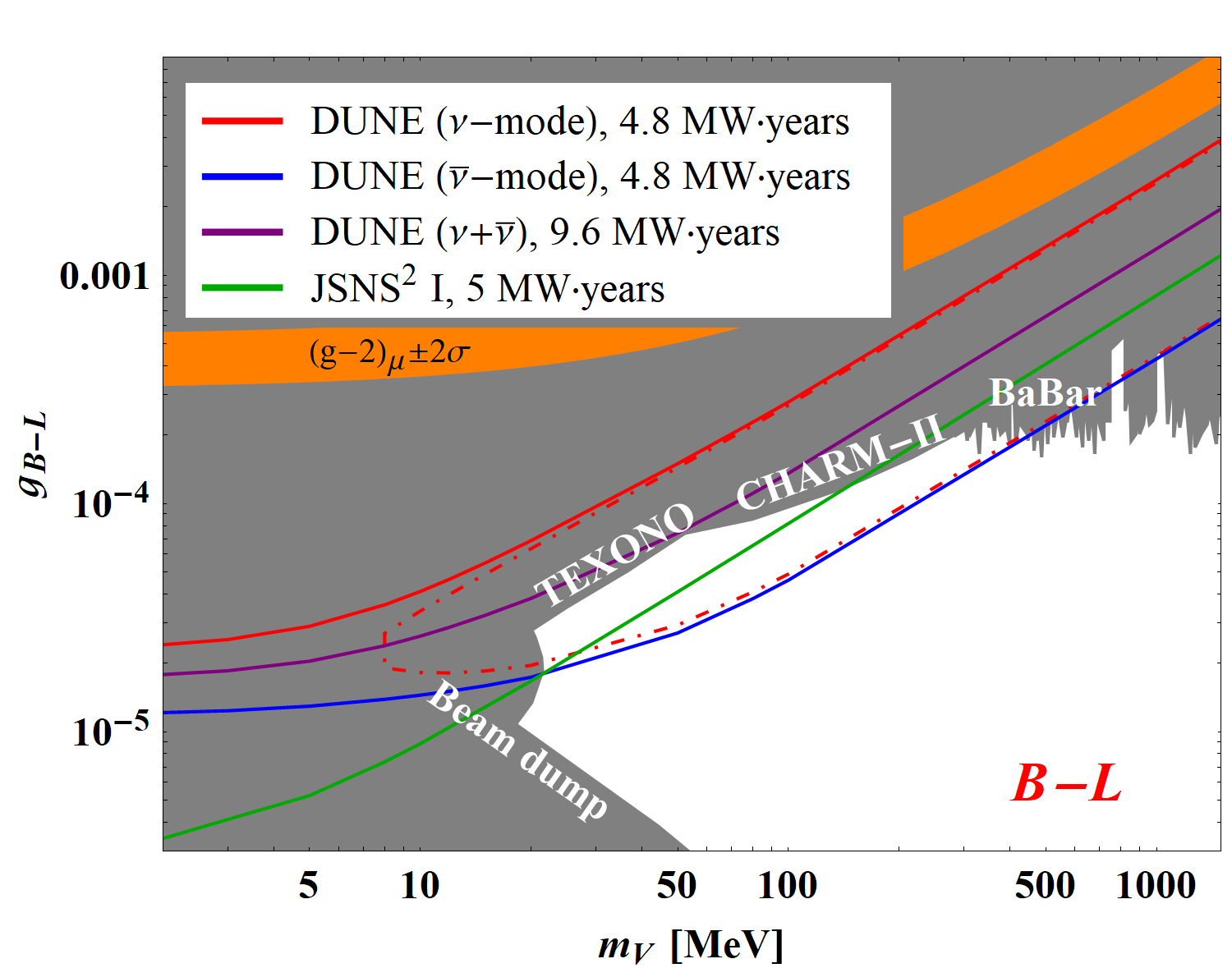}
     \includegraphics[width=8.9cm]{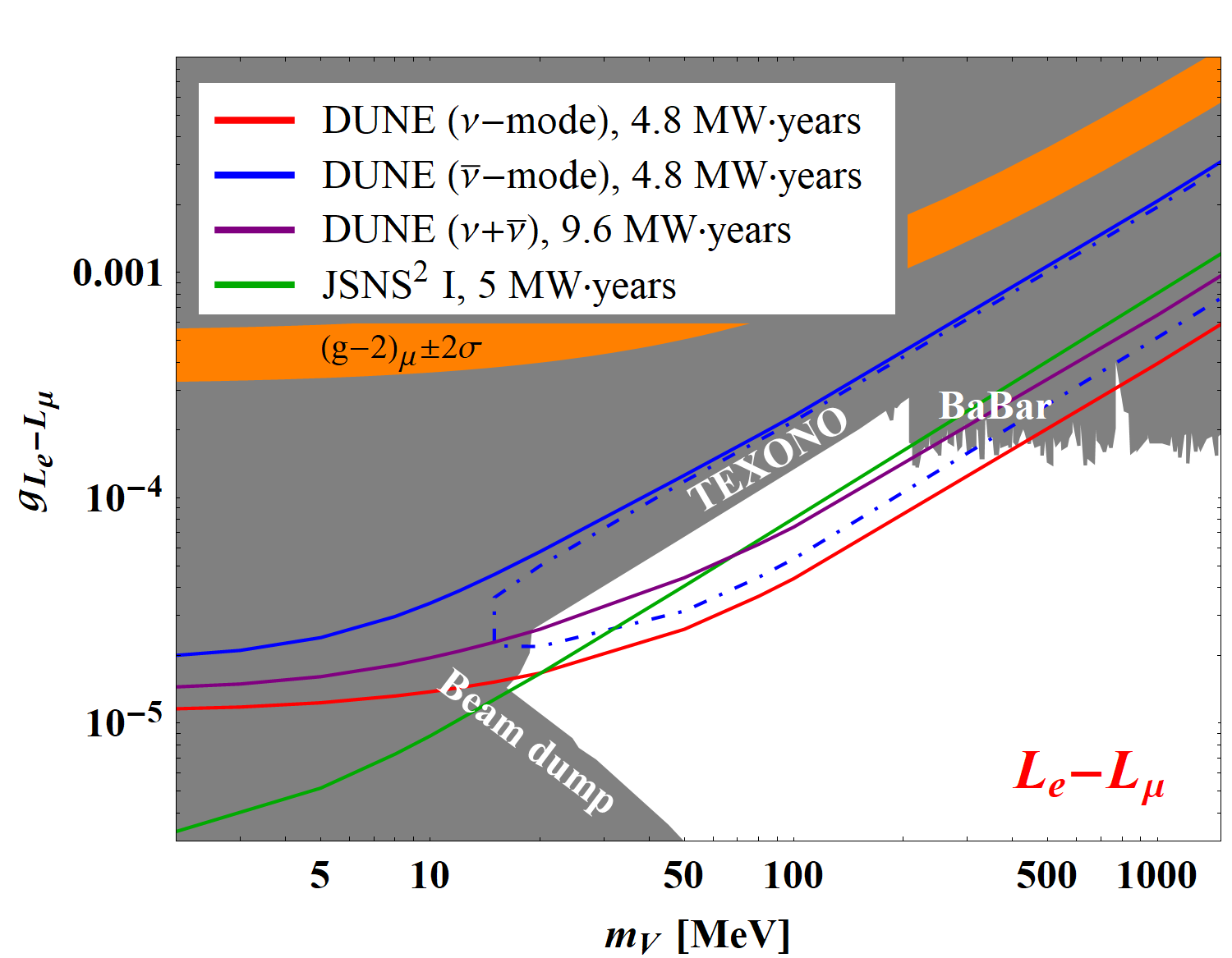}
    \caption{90\% C.L. sensitivity reaches of DUNE and JSNS$^2$ experiments in the plane of the $B-L$ (left panel) or $L_e-L_\mu$ (right panel) gauge boson mass $m_V$ and its coupling strength, under the assumption of the exposures shown in the legends. 
    We estimate the backgrounds using the number of neutrino-electron scattering events reported in Ref.~\cite{Marshall:2019vdy} for DUNE and calculating the pure SM expectation for JSNS$^2$.
    For DUNE, 5\% systematics in the neutrino flux normalization is included, while JSNS$^2$ is purely statistics-based. 
    The existing limits are shown by the gray-shaded region with experiments forming the boundaries explicitly mentioned in the figure.
    For comparison, the region preferred by the recently updated muon anomalous magnetic moment at $2\sigma$~\cite{Abi:2021gix} is shown as well. 
    DUNE can be sensitive to the signal by a deficit (the regions encompassed by dot-dashed lines) originating from a sizable destructive interference effect. See text for details.  }
    \label{fig:result}
\end{figure*}

\section{Data Analysis \label{sec:analysis}}
We now briefly describe how we perform our data analysis. 
For DUNE, we first estimate the expected number of events using the neutrino fluxes reported in Ref.~\cite{Abi:2020wmh}. 
We then assume a 67.5-ton detector fiducial mass and a 7-year exposure (3.5 years in the neutrino mode and 3.5 years in the antineutrino mode), with the energy threshold for recoiling electrons being 30 MeV~\cite{Abi:2020kei}. 
For the first six years, protons will be delivered by a 1.2-MW beamline, and then the beam power will be doubled~\cite{Abi:2020qib}. Therefore, the total statistics can be represented by 9.6 MW$\cdot$years, i.e., 4.8 MW$\cdot$years in each mode. 
Similarly, we make use of the neutrino flux reported in Ref.~\cite{Ajimura:2017fld} for JSNS$^2$, assuming a 17-ton fiducial mass and a 5-year exposure that are translated to statistics of 5 MW$\cdot$years, together with a detector threshold of 2.6 MeV~\cite{Ajimura:2017fld}. 

Since our signal is exactly the same as the usual $\nu$-$e^-$ scattering events except the mediator, any significant departure from the SM expectations (not only an excess but also a deficit) would herald the existence of a new gauge boson.  
For the DUNE-like experiments, there exists a dedicated study of estimating the $\nu$-$e^-$ scattering rate with nuclear effects included~\cite{Marshall:2019vdy}. We simply use the numbers reported therein to estimate the expected number of pure SM events, $N_{\rm SM}$, for DUNE.  
By contrast, a similar dedicated study for JSNS$^2$ is not yet available to the best of our knowledge. So, we assess $N_{\rm SM}$ for JSNS$^2$ using the cross-section of SM processes. 

One should be reminded that typical neutrinos at DUNE are energetic enough to induce charged-current quasi-elastic (CCQE) scattering events involving a nucleon recoil and an electron in the final state. 
Typical recoiling nucleons are not energetic enough to overcome the associated detector threshold, so that such CCQE events can mimic our signal events. 
It is well known that an $E_e\theta_e^2$ cut can significantly veto such CCQE-induced events thanks to the forwardness of the signal events under consideration (see e.g., Refs.~\cite{Park:2013dax,Park:2015eqa} for more details). 
On the other hand, CCQE events (almost) do not arise at JSNS$^2$ because neutrinos from the stopped pions and muons are not sufficiently energetic. 
Given all these considerations, we do not take such backgrounds into account in our analysis for simplicity.  

Finally, to estimate the sensitivity reaches of our benchmark experiments, we perform an unbinned $\chi^2$ analysis as our test statistic. Our $\chi^2$ is defined as follows: 
\begin{equation}
    \chi^2=\min_\alpha \left\{ \frac{[N_{{\rm SM}+V+{\rm int}}-(1+\alpha)N_{\rm SM}]^2}{N_{{\rm SM}+V+{\rm int}}}+\left(\frac{\alpha}{\sigma_{\rm norm}}\right)^2 \right\},
\end{equation}
where $N_{{\rm SM}+V+{\rm int}}$ is the total expected number of events including contributions of light gauge boson $V$. 
We also take into account the systematic uncertainty in the neutrino flux normalization along with the associated nuisance parameter $\alpha$. 
We consider a moderate 5\% uncertainty throughout our analysis for DUNE, while a 2\% level of systematics could be achieved~\cite{Abi:2020wmh}. But JSNS$^2$ will be purely statistics-based as the corresponding information is unavailable.  

\section{Results \label{sec:results}}
Our results are summarized in Fig.~\ref{fig:result} which shows the 90\% C.L. sensitivity reaches of DUNE and JSNS$^2$ experiments in the plane of the $B-L$ (left panel) or $L_e-L_\mu$ (right panel) gauge boson mass and its coupling strength. 
As described earlier, DUNE can collect the data in the neutrino and antineutrino modes separately, so we report not only the combined analysis but individual analyses. 
On the contrary, since JSNS$^2$ is incapable of taking the neutrino flux flavor-selectively, only the combined results are provided. 
For comparison, the existing limits compiled in e.g., Refs.~\cite{Ilten:2018crw, Bauer:2018onh, Dev:2021otb}, are shown by the gray-shaded regions with boundaries set by BaBar~\cite{BaBar:2014zli}, CHARM-II~\cite{CHARM-II:1993phx,CHARM-II:1994dzw,Bilmis:2015lja} (for the $B-L$ case), TEXONO~\cite{TEXONO:2002pra,TEXONO:2006xds,TEXONO:2009knm,Bilmis:2015lja,Lindner:2018kjo}, and beam-dump experiments (namely, E141~\cite{Riordan:1987aw}, Orsay~\cite{Davier:1989wz}, and U70/NuCal~\cite{Blumlein:2011mv,Blumlein:2013cua}).  The region preferred by the recently updated $(g-2)_\mu$ anomaly at 2$\sigma$~\cite{Abi:2021gix} is also included. 
Note that the leading limits by CHARM-II and TEXONO are from the neutrino-electron scattering, hence our results can be viewed as updating such limits.

A couple of points are worth being emphasized. Above all, our study suggests that DUNE should be sensitive to the signal by a {\it deficit} (the regions surrounded by dot-dashed lines) as well as by an excess, due to a sizable destructive interference effect.  
As discussed earlier, muon neutrinos (antineutrinos) in the neutrino (antineutrino) mode can induce substantial destructive interference if they have a gauge charge whose sign is equal (opposite) to that of the electron. 
Indeed, muon neutrinos (antineutrinos) dominate over the other flavors in the neutrino (antineutrino) mode~\cite{Abi:2020wmh}, and thus destructive interference can be prominent, not offset by constructive interference by other neutrino species. 
Also, the sensitivity of the $B-L$ and $L_e-L_\mu$ gauge bosons by a deficit arises in the neutrino and antineutrino modes, respectively, as predicted. 
If constructive interference only were considered, one of the two modes would show a much greater sensitivity than the other. But destructive interference allows for similar sensitivity reaches in a certain mass range, as suggested by Fig.~\ref{fig:result}. 
More importantly, our finding suggests that the combined analysis shown by purple curves\footnote{We have checked that the combined analysis for the $L_e-L_\mu$ case essentially yields sensitivity reaches similar to those in Ref.~\cite{Ballett:2019xoj}.} [$\chi^2 (\nu \hbox{ mode}+\bar\nu \hbox{ mode}$)] will not provide better results than the individual analyses [$\chi^2 (\nu \hbox{ mode}),~\chi^2(\bar\nu \hbox{ mode}$)]; the combined analysis would not allow DUNE to probe regions beyond the existing limits in the $B-L$ case.\footnote{Obviously, the sensitivity by $\chi^2 (\nu \hbox{ mode})+\chi^2(\bar\nu \hbox{ mode})$ can be better than that by $\chi^2 (\nu \hbox{ mode}),~\chi^2(\bar\nu \hbox{ mode}$), or $\chi^2 (\nu \hbox{ mode}+\bar\nu \hbox{ mode}$). }
This encourages previous beam-focused neutrino experiments to revisit their data and perform individual analyses upon the availability of mode-dependent data. 
By contrast, JSNS$^2$ does not show the sensitivity reaches by a deficit. The reason is that there is no way to distinguish neutrino and antineutrino fluxes and $\nu_\mu$ and $\bar{\nu}_\mu$ come with similar amounts, so that destructive interference is essentially washed out by constructive interference.   
Nevertheless, the higher beam intensity and the better detector proximity allow the {\it ongoing} JSNS$^2$ to achieve competitive sensitivity in the near future.   

Second, we see that our benchmark experiments can probe some regions of parameter space that are never been explored in any existing experiments or search efforts.
In addition, this indirect search channel can provide information complementary to direct search channels such as decay signals (e.g., Refs.~\cite{Bauer:2018onh,Ariga:2018uku,Berryman:2019dme,Dev:2021qjj}), as their best sensitivity coverage arises in different regions of parameter space. 

\section{Further Discussion \label{sec:discussion}} It is worth mentioning here that the main idea delineated thus far can be straightforwardly extended to the neutrino-nucleon/nucleus scattering processes, although the hadronic and nuclear uncertainties might weaken the detection prospects of the interference effect. Moreover, our idea is generically applicable to other flavor-selective experiments such as MINER$\nu$A, MiniBooNE, MicroBooNE, ICARUS, and SBND, as well as flavor non-selective experiments such as COHERENT and CCM. 
We reserve a detailed investigation of these intriguing possibilities for a follow-up study. 

The main idea proposed in this paper can also be generalized to other $U(1)'$ models; see Appendix.
As for the simplest  anomaly-free choices used here, we did not show here the results for other possibilities, such as the  $L_\mu-L_\tau$ model, because in this case, the light mediator does not couple to the electrons at leading order, so the limits from  neutrino-electron scattering will be weaker due to loop-suppression, as compared to other tree-level processes, such as the neutrino trident production~\cite{Ballett:2019xoj}. Similarly, for the $L_e-L_\tau$ case, the light mediator does not directly couple to $\nu_\mu$, thus we lose the advantage of the $\nu_\mu$ beam. Future proposals for an intense $\nu_\tau$ beam, such as DsTau~\cite{Aoki:2019jry}, can be useful in this context. 

\section{Conclusions \label{sec:conclusion}}
We have shown that destructive interference due to a light gauge boson mediator in neutrino-electron scattering could lead to a deficit in the neutrino (antineutrino) events of a particular flavor, depending on the mediator couplings. This will allow neutrino-flavor-selective experiments like DUNE to be able to probe a larger parameter space, while running in the neutrino (antineutrino) mode alone, as compared to the combined neutrino and antineutrino run. This is largely complementary to other current and future searches for light gauge bosons, such as those involving its production and  decay in beam-dump and collider facilities.  Moreover, since the destructive interference effect is prominent only in either neutrino or antineutrino mode, depending on the mediator type, it provides a new way to distinguish between different mediators, should such an deficit arise in the experimental data.

\acknowledgements 
We would like to thank Zahra Tabrizi for useful discussions and clarification regarding the results of Ref.~\cite{Ballett:2019xoj}. We also thank Peter Denton for discussion on the constraints from neutrino experiments. The work of BD is supported in part by the US Department of Energy under Grant No.~DE-SC0017987, by the Neutrino Theory Network Program, and by a Fermilab Intensity Frontier Fellowship. The work of DK is supported by DOE under Grant No. DE-SC0010813. The work of KS is supported by U.S. Department of Energy grant number~DE-SC0009956.
The work of YZ is supported  by  the  Fundamental Research Funds for the Central Universities.

\section*{Appendix} 

\noindent {\bf Generalization to other $U(1)'$ models}.
In the most general case, the associated $Z'$ gague boson interactions with charged leptons and neutrinos can be written as (with $f = \nu,\, \ell$), 
\begin{eqnarray}
-{\cal L} \ \supset \ Z'_{\mu} \left( g_{V\alpha} \bar{f}_\alpha \gamma^\mu f_\alpha + g_{A\alpha} \bar{f}_\alpha \gamma^\mu \gamma_5 f_\alpha \right) \,, 
\end{eqnarray}
where $\alpha = e\;\mu,\;\tau$ is the lepton flavor index, and gauge charges are omitted for brevity.
In this context, the corresponding analytic expressions for the pure $Z'$ and interference terms can be obtained as follows:  
\begin{widetext} 
\begin{align}
\frac{d}{dE_e} \sigma_{Z'} \left(
 \overset{\scriptscriptstyle(-)}{\nu}_\alpha e^- \to
  \overset{\scriptscriptstyle(-)}{\nu}_\alpha e^- \right) 
 & \ = \  \frac{m_e g^2_{-\alpha}}{4\pi  (m_{Z^\prime}^2 + 2 m_e E_e)^2} F_1 \,, \\
\frac{d}{dE_e} \sigma_{\rm int} \left(
 \overset{\scriptscriptstyle(-)}{\nu}_\alpha e^- \to
  \overset{\scriptscriptstyle(-)}{\nu}_\alpha e^- \right) %\nonumber \\ 
& \ = \ \frac{m_e g_{-\alpha} G_F}{2 \sqrt{2}\pi  (m_{Z^\prime}^2 + 2 m_e E_e)}
\big( F_2 + s_W^2  F_3 \big) \,, 
\end{align}
\end{widetext}
where the dimensionless parameters $F_{1,2,3}$ are given by
\begin{align}
F_1 & \ = \ 2 (g_{Ve}^2 + g_{Ae}^2) + g_{\pm e}^2 x^2 - 2 g_{\pm e}^2 x - g_{+e}g_{-e} xx_0 \, , \\
F_{2} & \ = \  S_{\alpha e} \left( 2g_{-e} - g_{+e} xx_0 \right) -
2\delta S_{\alpha e} g_{-e} x  \left( 2 - x \right) \, ,  \\ 
F_{3} & \ = \ -  4g_{\pm e} x \left( 2-x \right) + 4 g_{Ve} \left( 2 - xx_0 \right) \,,
\end{align}
with $g_{\pm \alpha} = g_{V\alpha} \pm g_{A\alpha}$, $x={E_e}/{E_\nu}$, $x_0 = {m_e}/{E_\nu}$, $S_{\alpha e} = 1$ for $\alpha = e$ and $-1$ for $\alpha =\mu,\,\tau$, $\delta = 0$ for all three flavors of neutrinos and $1$ for all antineutrinos, $g_{\pm\alpha} = g_{+\alpha}$ for neutrinos and $g_{-\alpha}$ for antineutrinos. It is straightforward to show that the choices of $g_{Ve}=Q_e g_V$, $g_{V\alpha}=Q_{\nu_\ell}g_V$, $g_{A\alpha}=0$, and $m_{Z'}=m_V$ can reproduce Eqs.~\eqref{eq:zpcont} and \eqref{eq:intcont}. 
A detailed phenomenological study of neutrino-electron/nucleon scattering for different choices of $g_{V\alpha}$ and $g_{A\alpha}$ is deferred to a future study.

\bibliography{main}

\end{document}